\documentclass[aps,nofootinbib]{revtex4} 
\usepackage{graphicx}
\input epsf
\newcommand{\sfig}[2]{
\includegraphics[width=#2]{#1}
        }
\newcommand{\Sfig}[2]{
    \begin{figure}[thbp]
    \sfig{#1.eps}{0.9\columnwidth}
    \caption{{\small #2}}
    \label{fig:#1}
    \end{figure}
}

\newcommand{\rf}[1]{\ref{fig:#1}}

\def\vl{{\bf l}}
\def\vt{{\bf\theta}}
\def\vx{{\bf x}}
\def\rs{\rm rs}
\def\lle{\rm ll}
\def\born{\rm Born}

\def\cmm2{{\,\rm cm^{-2}}}
\def\cm2{{\,{\rm cm}^2}}
\def\cmm3{{\,{\rm cm}^{-3}}}
\def\gcmm3{{\,{\rm g\,cm^{-3}}}}

\def\fun#1#2{\lower3.6pt\vbox{\baselineskip0pt\lineskip.9pt
  \ialign{$\mathsurround=0pt#1\hfil##\hfil$\crcr#2\crcr\sim\crcr}}}

\def\be{\begin{equation}}
\def\ee{\end{equation}}
\def\bea{\begin{eqnarray}}
\def\eea{\end{eqnarray}}
\newcommand{\vs}{\nonumber\\}

\newcommand{\ec}[1]{Eq.~(\ref{eq:#1})}

\newcommand{\eql}[1]{\label{eq:#1}}

\begin{document}

\title{The Weak Lensing Bispectrum}

\author{Scott Dodelson$^{1,2,3}$ and Pengjie Zhang$^1$}

\affiliation{$^1$NASA/Fermilab Astrophysics Center
Fermi National Accelerator Laboratory, Batavia, IL~~60510-0500}
\affiliation{$^2$Department of Astronomy \& Astrophysics, The University of Chicago,
Chicago, IL~~60637-1433}
\affiliation{$^3$Department of Physics and Astronomy, Northwestern University,
Evanston, IL~~60208}

\date{\today}
\begin{abstract}
Weak gravitational lensing of background galaxies offers an
excellent opportunity to study the intervening distribution of
matter. While much attention to date has focused on the two-point
function of the cosmic shear, the three-point function, the {\it
bispectrum}, also contains very useful cosmological information.
Here, we compute three corrections to the bispectrum which are
nominally of the same order as the leading term. We show that the
corrections are small, so they can be ignored when analyzing present
surveys. However, they will eventually have to be included for
accurate parameter estimates from future surveys.
\end{abstract}
\maketitle

\section{Introduction}

Weak lensing offers cosmologists the opportunity to probe the
distribution of mass in the universe~\cite{Bartelmann:1999yn}. This
prospect is so alluring because theories make first-principles
predictions about this distribution, so we can hope to extract
important constraints on fundamental cosmological parameters from
weak lensing surveys~\cite{Benabed:2001dm,Hu:2001fb,Huterer:2001yu,
Abazajian:2002ck,Refregier:2003xe,Simon:2003ws,
Bernstein:2003es,Heavens:2003jx,Takada:2003ef}.
In many senses, this promise is similar to that felt by those who
studied the cosmic microwave background (CMB) a decade ago:
theoretical predictions are straightforward; experiments have
detected the effect (anisotropies in that case and cosmic shear in
this); and there are grand plans for the future (which have been
realized in the case of the CMB).

Armed with this optimism, cosmologists are quick to throw in warning labels: the signal
is extremely small, so the only way to measure cosmic shear is to average over many
background galaxies. Each individual galaxy is observed with its own set of systematics (seeing,
elliptical point spread functions, calibration, unknown or at least uncertain
redshift, etc.) and these vary from one galaxy to
another. Measurements of cosmic shear are unlikely to produce smooth maps since
there inevitably will be bright stars which must be masked out. Accounting for these
masks leads to complicated window functions. It is not clear then that weak lensing measurements
will eventually pay off as did those of the CMB.

There is one area though in which weak lensing measurements have an
advantage over the CMB: the higher point functions of the cosmic
shear field are potentially simpler to interpret and more relevant
than those in the CMB. Naively, this is what one would expect, for
the cosmic shear field is sensitive to the matter density which has
gone nonlinear and therefore will have large corrections to the
Gaussian limit. Temperature fluctuations on the other hand are still
stuck at the $10^{-5}$ level, so are expected to be very close to
Gaussian (recall that the initial distribution of both temperature
and matter inhomogeneities was likely Gaussian and for a Gaussian
distribution the higher point functions are trivially related to the
two-point function). We might expect then the 3-point function of
CMB anisotropies to be very small, while that of the cosmic shear to
be quite large, at least on small scales. Much work has been done
over the past few years attempting to debunk these naive ideas. We
have found that the higher point functions of the CMB are quite
interesting:
lensing~\cite{Zaldarriaga:2000ud,Zaldarriaga:1998te,Hu:2001tn}, hot
gas~\cite{SZ}, peculiar velocities~\cite{SZv}, and
reionization~\cite{Cooray:1999kg,castro} leave their imprint in
these higher point functions. Nonetheless, the fact remains that to
date there has been no detection of a non-zero 3-point function, for
example, in the CMB, while several
groups~\cite{Bernardeau02,Pen:2003vw} have claimed such a detection
in the cosmic shear field. Further, a number of
authors~\cite{Hui99,Benabed:2001dm,Cooray:2000uu,Refregier:2003xe,Takada:2003ef}
have showed that the bispectrum of the cosmic shear field will be
able to constrain important cosmological parameters, including
properties of the dark energy. Certainly, then, we need to obtain
accurate predictions of the bispectrum of the shear.

With this in mind, it is important to emphasize just how important
``higher order'' corrections to the bispectrum might be. To
understand this, recall that, to lowest order, the shear is a
line-of-sight integral over the matter overdensity: \be \gamma({\bf
\theta}) = \int d\chi f(\chi) \delta({\bf x}(\chi,{\bf\theta}))
\eql{Igamdel}\ee where ${\bf\theta}$ is the angular position on the
sky, $f$ is a weighting function, and $\chi$ is the comoving
distance along the line of sight. Roughly then the bispectrum, which
is proportional to $\langle \gamma^3 \rangle$, is proportional to
$\langle\delta^3\rangle$. It vanishes therefore in the large-scale
limit where the density field is Gaussian and nonlinearities are
irrelevant. The main contribution to the bispectrum then comes from
the fact that, due to gravity, $\delta$ evolves nonlinearly. In
perturbation theory, we would write $\delta=\delta^{(1)} +
\delta^{(2)} + \ldots$ where $\delta^{(1)}$ is the linear
overdensity and $\delta^{(2)}$ is proportional to $\delta_L^2$. This
main contribution to the bispectrum then comes from terms
proportional to $\langle \delta^{(1)} \delta^{(1)}
\delta^{(2)}\rangle $ and so is proportional to $\delta_L^4$.

It is clear then that any correction which alters the linear
relation between shear and overdensity in \ec{Igamdel} is of the
same order in $\delta_L$ as the ``main contribution.'' Here we study
three such corrections first identified by Schneider et
al.~\cite{Schneider:1997ge} and compute their effect on the
bispectrum:
\begin{itemize}
\item {\bf Reduced Shear} We estimate shear by measuring ellipticities of background
galaxies, invoking the relation~\cite{Bartelmann:1999yn,Dodelson} $\epsilon_i=2\gamma_i$,
where the subscript $i$ refers to the two components of ellipticity/shear. This relation
though is only approximate; the full relation is
\be
\epsilon_i = {2\gamma_i\over 1-\kappa}
\ee
where $\kappa$ is the convergence. Expanding the denominator, we see that the ellipticities used
to estimate cosmic shear have terms quadratic in the perturbations ($2\gamma_i\kappa$). This quadratic
term contributes to the bispectrum at the same order as the
main term.
\item {\bf Lens-Lens Coupling} \ec{Igamdel} does not account for the fact that lenses are correlated along
the line of sight. This lens-lens coupling induces another quadratic term in the relation between shear and
overdensity.

\item {\bf Born Approximation} When computing the shear one
integrates along the photon path back towards the source. There is
a complication inherent in this integration encoded in the
argument of the overdensity in \ec{Igamdel}: what is the position
of the photon at radial distance $\chi$ if its observed angular
position today is ${\bf\theta}$? the naive answer ${\bf
x}^{(0)}=\chi({\bf\theta},1)$ is the position corresponding to the
path taken by an undeflected photon. Expanding about this zero
order position leads to a correction proportional to $\nabla
\delta \cdot [{\bf x}-{\bf x}^{(0)}]$. Since ${\bf x}$ differs
from the undeflected position only if $\delta$ is nonzero, this
correction is second order in $\delta$. It too contributes a term
of order $\delta^4$ to the bispectrum.
\end{itemize}

The next section reviews some basic lensing results including the standard computations of the
power spectrum and the bispectrum. \S{III} computes the correction to the bispectrum from
the three effects enumerated above. The bispectrum cannot be simply plotted on a 2D graph
since it depends on the three variables required to specify a triangle. Therefore, \S{IV}
examines various ways of condensing the information contained in these corrections. The goal is
to see whether the corrections are important.

\section{Review of Basic Results}

The deformation tensor is defined as the deviation from unity of the Jacobian relating
the undeflected position (${\bf\theta_S}$) to the actual position (${\bf\theta}$):
\be
\psi_{ij} \equiv  \delta_{ij} - {\partial \theta_{S,i} \over \partial\theta_j}
.\ee
The elements of this $2\times2$ matrix are the two components of shear and the convergence:
\be
\psi_{ij} \equiv \left(  \matrix{ \kappa + \gamma_1 & \gamma_2\cr
            \gamma_2 & \kappa -\gamma_1} \right)
.\eql{def}\ee
These are simply definitions. The physics comes from
solving the geodesic equation and expressing the distortion tensor
in terms of the gravitational potential $\Phi$: \be
\psi_{ij}(\vt,\chi_s)
 =  \int_0^{\chi_s} d\chi W(\chi,\chi_s){\partial^2\Phi\left(\vx(\vt,\chi);\chi\right)\over \partial x_i\partial x_k}
 \left[ \delta_{kj} - \psi_{kj}(\vt,\chi)\right]
.\eql{psi}\ee
Here, we are assuming a flat universe;
sources are assumed to lie at $\chi_s$; the first argument of the $\Phi$ is
the 3D comoving position (in the small angle limit)
\be
\vx(\vt,\chi)= [\chi\theta_x,\chi\theta_y,\chi] - \int_0^{\chi} d\chi' W(\chi',\chi) {\chi\over \chi'}
\nabla\Phi\left(\vx(\vt,\chi');\chi\right)
\eql{defx}\ee
while the second ($\chi$) refers to the cosmic time at which the photon path passed by this position.
The weighting function is
\be
W(\chi,\chi_s) = 2 \chi (1-\chi/\chi_s) \Theta(\chi_s-\chi)
.\ee
We observe ellipticities of background galaxies $\epsilon_i$, which are related to
the elements of the distortion tensor via
\be
\epsilon_i = {2\gamma_i\over 1-\kappa}.
\ee

One way to estimate the convergence, which is the projected
density, is to work in Fourier space; then, \be
\hat{\tilde\kappa}(\vl) =  {1\over l^2} T_i(\vl)
\tilde\epsilon_i(\vl) .\eql{kest}\ee Here the variable conjugate
to ${\bf \theta}$ is $\vl$. As usual, large $l$ corresponds to
small scales. The trigonometric functions in \ec{kest} are defined
with respect to an arbitrary $x$- axis as \bea T_1(\vl) &\equiv &
{l_x^2 - l_y^2\over 2} = {l^2\over 2} \cos(2\phi_l)\vs T_2(\vl)
&\equiv & l_xl_y  = {l^2\over 2} \sin(2\phi_l) .\eql{deft}\eea
Here $\phi_l$ is the angle between $\vl$ and the $x$- axis. In the
limit in which $\epsilon_i\rightarrow 2\gamma_i$,  the estimator
in \ec{kest} reduces to \bea \hat{\tilde\kappa}(\vl)
&\overbrace{\longrightarrow}^{\rm reduced\ shear}&  \cos(2\phi_l)
\tilde\gamma_1(\vl) + \sin(2\phi_l) \tilde\gamma_2(\vl) \vs &=&
\cos(2\phi_l) (\tilde\psi_{11}(\vl) - \tilde\psi_{22}(\vl))/2 +
\sin(2\phi_l) \tilde\psi_{12}(\vl) \vs
&\overbrace{\longrightarrow}^{\rm lens-lens}& {-l^2\over 2}
\tilde\psi(\vl) .\eql{psil}\eea The approximation on the first
line neglects the fact that ellipticities are sensitive to the
reduced shear; the approximation on the third line neglects the
second order term in the brackets in \ec{psi}, the term which
accounts for the fact that the lens distribution is correlated.
The projected potential in \ec{psil} is defined as
\newcommand{\psiz}{\psi^{(0)}}
\be
\psi(\vt,\chi_s)
 =  \int_0^{\chi_s} d\chi {W(\chi,\chi_s)\over \chi^2} \Phi\left(\vx(\vt,\chi);\chi\right)
.\ee
In the Born approximation the gravitational potential in the integral along the line of sight is evaluated
at the {\it unperturbed} path. In that case, its Fourier transform reduces to
\be
\tilde\psi(\vl,\chi_s) \overbrace{\longrightarrow}^{\rm Born} \tilde\psiz(\vl,\chi_s) =
\int_0^{\chi_s} d\chi {W(\chi,\chi_s)\over \chi^2} \tilde\phi\left(\vl;\chi\right)
\eql{bornapp}\ee
with
\be
\tilde\phi(\vl,\chi)\equiv {1\over \chi^2} \int {dk_3\over 2\pi}
\tilde\Phi(\vl/\chi,k_3;\chi) e^{ik_3\chi}
.\ee
Note that $\tilde\phi$ is dimensionless unlike $\tilde\Phi$ which has dimensions
of (length)$^3$.

The statistics of $\tilde\psiz$ follow directly from those of
$\tilde\phi$, which are simple if we include only modes with $k_3$
small, i.e. the Limber approximation~\cite{Limber, Dodelson},
 \be
\langle\tilde\phi(\vl,\chi) \tilde\phi(\vl',\chi')\rangle
=(2\pi)^2\delta^2(\vl+\vl')\delta(\chi-\chi')P_\Phi(l/\chi;\chi)/\chi^2
.\eql{statphi}\ee Here $P_\Phi$ is the power spectrum of the
gravitational potential. Similarly, the three point-function is
related to the spatial bispectrum~\cite{castro}: \be
\langle\tilde\phi(\vl_1,\chi_1)
\tilde\phi(\vl_2,\chi_2)\tilde\phi(\vl_3,\chi_3)\rangle
=(2\pi)^2\delta^2(\vl_1+\vl_2+\vl_3)
\delta(\chi-\chi')\delta(\chi-\chi'')B_\Phi(\vl_1/\chi,\vl_2/\chi,\vl_3/\chi;\chi)/\chi^4
.\ee Then, we have \be \langle \tilde\psiz(\vl)
\tilde\psiz(\vl')\rangle = (2\pi)^2 \delta^2(\vl+\vl') P_2(l)
,\eql{2point}\ee with \be P_2(l)=  \int_0^\infty d\chi
{W^2(\chi,\chi_s)\over \chi^6} P_\Phi(l/\chi;\chi) .\eql{p2}\ee
Similarly, the three-point function is \be \langle
\tilde\psiz(\vl_1) \tilde\psiz(\vl_2)\tilde\psiz(\vl_3)\rangle =
(2\pi)^2 \delta^2(\vl_1+\vl_2+\vl_3) P_3(\vl_1,\vl_2,\vl_3) ,\ee
where now the projected power is a line-of-sight integral over the
bispectrum: \be P_3(\vl_1,\vl_2,\vl_3) = \int_0^\infty d\chi
{W^3(\chi)\over \chi^{10}}
B_\Phi(\vl_1/\chi,\vl_2/\chi,\vl_3/\chi;\chi) .\ee

The power spectrum, or the $C_l$'s, are defined as the coefficient
of $(2\pi)^2\delta(\vl+\vl')$ when computing the variance of
$\hat{\tilde\kappa}$. Since $\hat{\tilde\kappa}=-l^2\psiz/2$ in the
standard computation, we have \be C^\kappa_l=l^4P_2(l)/4.\ee
Similarly, the bispectrum of the $\kappa$\ estimator is \be
B^\kappa(\vl_1,\vl_2,\vl_3) = -l^6P_3(\vl_1,\vl_2,\vl_3)/8,\ee in
agreement\footnote{One subtlety when comparing with other results is
the sign. The sign here is negative because $\tilde\Phi\propto
-\tilde\delta$.} with previous results~\cite{Takada:2003ef}. The
bispectrum with all $l$'s equal, the equilateral configuration, is
shown in Fig.~\rf{zero}.

\Sfig{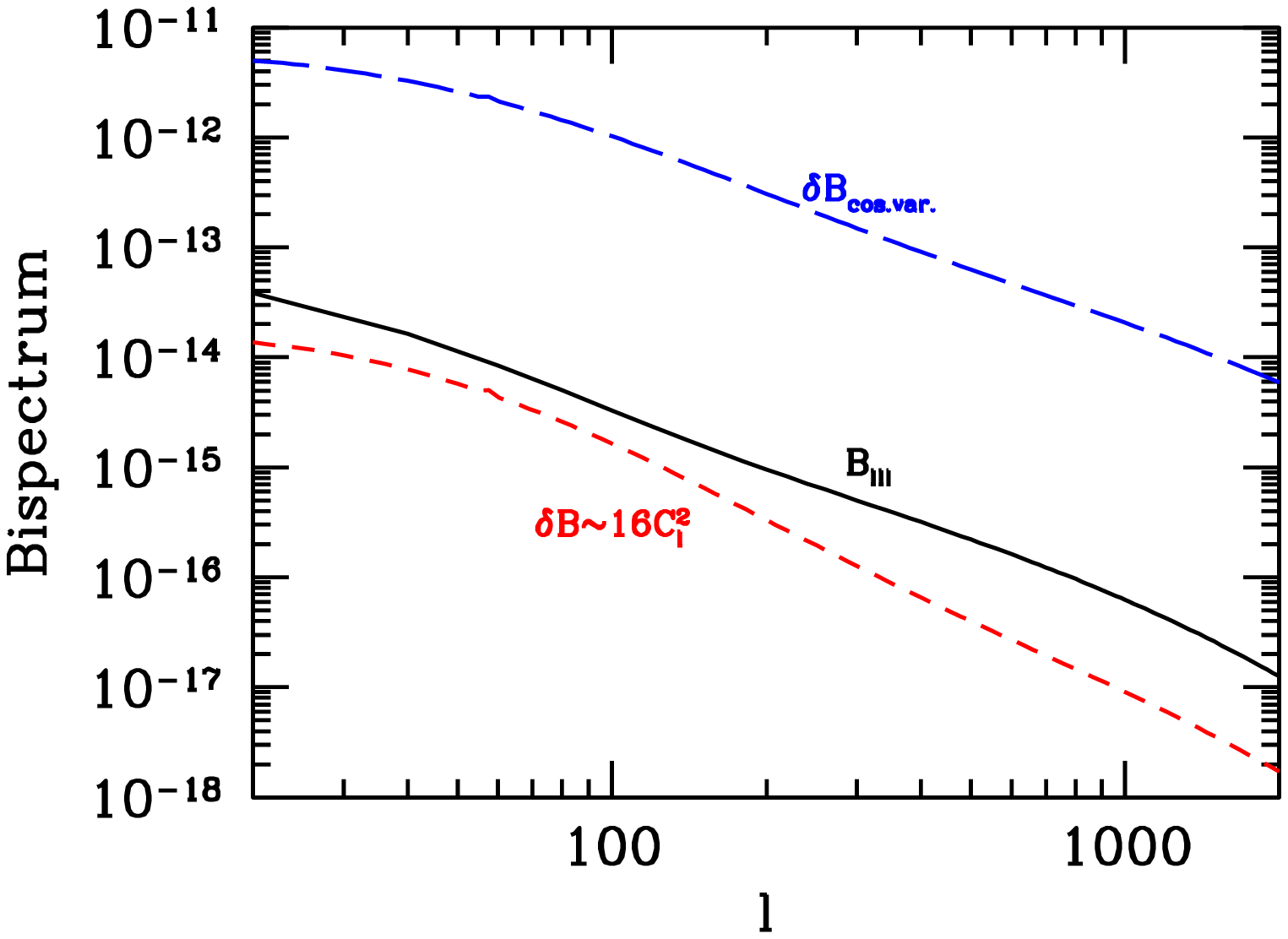}{Equilateral bispectrum. Solid curve is the standard result; short-dashed [red] curve
is order of magnitude of corrections considered here; long-dashed [blue] curve is cosmic-variance
error. The signal to noise from a single configuration is therefore extremely small.}

A few qualitative comments are in order here. The standard measure
of the amplitude of fluctuations is $l^2 C_l^{\kappa} = l^6 P_2/4$.
Let's do an order of magnitude estimate for this quantity $l^6 P_2$
in terms of the amplitude of density fluctuations, $\Delta^2 \equiv
k^3 P_\delta/2\pi^2$. Since $l\sim (k/H_0)$ and since $\Phi\sim
(H_0/k)^2 \delta$, we have $P_\Phi \sim P_\delta/l^4$. Now, \ec{p2}
suggests that $P_2\sim P_\Phi/\chi^3$; since $\chi\sim l/k$,
$P_2\sim (k/l)^3 P_\Phi \sim (k/l)^3 P_\delta /l^4 \sim
\Delta^2/l^7$. We expect then that $l^6P_2$ should be of order
$\Delta^2/l$. What this means physically is that projection effects
suppress the 2D power spectrum by a factor of $1/l$. There is even a
nice explanation of this in terms of Fourier
modes~\cite{Kaiser:1991qi,Dodelson}: only modes with small $k_3$
contribute; these are a fraction of $1/l$ of the total number of
modes. The bottom line then is that the power spectrum of the
convergence is smaller than the power spectrum of the 3D density
field.

Similar order of magnitude estimates relate the angular bispectrum, $-l^6P_3/8$,
to the 3D bispectrum of the density field:
\be
 l^6P_3 \sim l^6 \left[(k/l)^6 B_\Phi\right]
 \sim H_0^6 B_\delta
 .\ee
The corrections we consider below are all of order $l^8P_2^2$, which
by the arguments of the preceding paragraph are of order $H_0^6 P_\delta^2$.
The 3D bispectrum $B_\delta$ is nominally of the same order as the square of
the power spectrum, $P_\delta^2$. However, numerically it is a bit larger~\cite{Bernardeau:2001qr},
as indicated in Fig.~\rf{zero},
so the corrections we compute are not as important as we would have hoped.

\section{Higher Order Terms}

We now compute the corrections to the bispectrum from going beyond
the approximations in \ec{psil} and \ec{bornapp}.

\subsection{Reduced Shear}

The first-order correction to the $\epsilon-\gamma$ relation is
\be \epsilon_i^{\rm rs} = 2\gamma_i\kappa. \ee When we switch to
Fourier space, the relation between ellipticity and shear is a
convolution integral (products in real space correspond to
convolutions in Fourier space): \be \tilde\epsilon_i^{\rm rs}(\vl)
= \int {d^2l'\over (2\pi)^2} l'^2 T_i(\vl-\vl')
\tilde\psiz(\vl')\tilde\psiz(\vl-\vl') \eql{eprs}\ee When we form
the bispectrum  of the $\kappa$ estimator (\ec{kest}), the first
order correction emerges by replacing one of the three
ellipticities with the higher order \ec{eprs}. So this correction
to the bispectrum estimator becomes:
\bea \langle
\hat{\tilde\kappa}(\vl_1) \hat{\tilde\kappa}(\vl_2)
\hat{\tilde\kappa}(\vl_3) \rangle^{\rs} &=& {T_i(\vl_1)
l_2^2l_3^2\over 4l_1^2} \int {d^2l'\over (2\pi)^2} l'^2
T_i(\vl_1-\vl') \langle
\tilde\psiz(\vl')\tilde\psiz(\vl_1-\vl') \tilde\psiz(\vl_2)
\tilde\psiz(\vl_3) \rangle \vs &&+ \left( \vl_1\leftrightarrow
\vl_2 \right) + \left( \vl_1\leftrightarrow \vl_3 \right)
.\eql{hatll}\eea

The four-point function for the potential gets contributions from
the connected part -- the trispectrum -- and the disconnected
part: the product of power spectra. Here we consider only the
latter set of terms as these are expected to dominate. That is,
let
\bea \langle \tilde\psiz(\vl')\tilde\psiz(\vl_1-\vl')
\tilde\psiz(\vl_2) \tilde\psiz(\vl_3) \rangle &\rightarrow &(2\pi)^4
P_2(\vl_2) P_2(\vl_3)
 \vs
&&\times \Big[ \delta^2(\vl'+\vl_2) \delta^2(\vl_1-\vl'+\vl_3) +
\delta^2(\vl'+\vl_3) \delta^2(\vl_1-\vl'+\vl_2) \Big].\vs
 \eea
 The
integral over $\vl'$ then leaves the coefficient of
$(2\pi)^2\delta^2(\vl_1+\vl_2+\vl_3)$ as \be
 {T_i(\vl_1) l_2^2 P_2(l_2) l_3^2 P_2(l_3) \over 4l_1^2}
\left[ l_2^2 T_i(\vl_3) + l_3^2 T_i(\vl_2) \right] \ee plus
permutations. The contraction over the geometric factors involves
\be T_i(\vl_1) T_i(\vl_2) = {l_1^2 l_2^2\over 4} \cos(2\phi_{12})
\ee where $\phi_{12}$ is the angle between $\vl_1$ and $\vl_2$.

Defining $B_{123}\equiv B(\vl_1,\vl_2,\vl_3)$ as the coefficient
multiplying $(2\pi)^2\delta^2(\vl_1+\vl_2+\vl_3)$, we therefore
have
\be B_{123}^{\rs} =  {l_2^4l_3^4 P_2(l_2)P_2(l_3)\over
16}\left[ \cos(2\phi_{13}) + \cos(2\phi_{12})\right]\eql{rsbi} \ee
plus
permutations.

\subsection{Lens-Lens Coupling}

Lens-lens coupling in encapsulated by the second term in square
brackets in \ec{def}. This contribution to the distortion tensor
is then \be \psi^{\lle}_{ab}(\vt,\chi_s) = -\int d\chi
W(\chi,\chi_s) \Phi_{,ac} \psi_{cb}(\vt,\chi). \ee At second order
in $\Phi$, this reduces to
\be \psi^{\lle}_{ab}({\vt},\chi_s)=-
\int d\chi W(\chi,\chi_s)\Phi_{,ac}(\vt,\chi)\int
d\chi' W(\chi',\chi)  \Phi_{,cb}(\vt,\chi') .\ee

Using \ec{kest}, we can compute how this term contributes to the
estimator of convergence. The second-order contribution is
\bea \hat\kappa^{\lle}(\vl) &=& {2 T_i(\vl)
\gamma_i^{\lle}(\vl)\over l^2} \vs &=& {-T_i(\vl)\over l^2}\int
{d^2l'\over (2\pi)^2} E_i(\vl',\vl-\vl') \int {d\chi\over \chi^2}
W(\chi,\chi_s) \int {d\chi'\over \chi'^2}
W(\chi',\chi) \tilde\phi(\vl',\chi) \tilde\phi(\vl-\vl',\chi')
.\eea
Here the geometrical factors are defined as
\bea
E_1(\vl_1,\vl_2) &\equiv &\vl_1\cdot\vl_2 \left[ \vl_{1,x} \vl_{2,x} - \vl_{1,y} \vl_{2,y}\right]
\vs
E_2(\vl_1,\vl_2) &\equiv &\vl_1\cdot\vl_2 \left[ \vl_{1,x} \vl_{2,y} + \vl_{1,y}
\vl_{2,x}\right]
.\eea

The estimator for the bispectrum of the convergence then gets a Gaussian
contribution from the lens-lens term. One such term is
\begin{widetext}
\bea
\langle \hat\kappa^{\lle}(\vl_1) \hat\kappa^{(0)}(\vl_2)
\hat\kappa^{(0)}(\vl_3) \rangle
&=& {- l_2^2l_3^2 T_i(\vl_1)\over 4l_1^2} \int {d\chi\over \chi^2} W(\chi,\chi_s)
\int {d\chi'\over\chi'^2} W(\chi',\chi) \int {d\chi_2\over \chi_2^2}
W(\chi_2,\chi_s)
\int {d\chi_3\over\chi_3^2} W(\chi_3,\chi_s)  \vs
&&\times\int {d^2l'\over (2\pi)^2} E_i(\vl',\vl_1-\vl')
\langle \tilde\phi(\vl',\chi) \tilde\phi(\vl_1-\vl',\chi')
\tilde\phi(\vl_2,\chi_2) \tilde\phi(\vl_3,\chi_3) \rangle
.\eea
\end{widetext}
Two other terms exist with $\vl_1\leftrightarrow\vl_2$ and
$\vl_1\leftrightarrow\vl_3$. In this case, the trispectrum does not
contribute in the Limber approximation. Physically, the Limber
approximation sets all lenses close to each other; mathematically,
this corresponds to enforcing the constraint that the line of sight
distances are all equal~\cite{castro}. Here this constraint sets
$\chi=\chi'=\chi_2=\chi_3$, so that $W(\chi',\chi)$ vanishes. The
only relevant terms therefore are the two products of two-point
functions. Momentum conservation from the first such pair enforces
$\vl'=-\vl_2$ and $\vl_1-\vl'=-\vl_3$. Each two point function is
evaluated using the Limber approximation as in \ec{statphi}.

The bispectrum from lens-lens coupling is then \bea B_{123}^{\lle}
&=& {-l_2^2l_3^2 T_i(\vl_1)E_i(\vl_2,\vl_3)\over 4l_1^2}
 \int {d\chi\over \chi^6} W^2(\chi,\chi_s)
\int {d\chi'\over\chi'^6} W(\chi',\chi) W(\chi',\chi_s)   \vs
&&\times\left[ P_\Phi(l_2/\chi;\chi) P_\Phi(l_3/\chi';\chi')
+ P_\Phi(l_3/\chi;\chi) P_\Phi(l_2/\chi';\chi')
\right] + (\vl_1\leftrightarrow\vl_2) +
(\vl_1\leftrightarrow\vl_3).\eql{billint}\eea
Here we have used the fact that $E_i(\vl_1,-\vl_2)=E_i(\vl_1,\vl_2)
=E(\vl_2,\vl_1)$.

The geometrical factor in front reduces to \be
T_i(\vl_1)E_i(\vl_2,\vl_3) =
{l_1^2l_2^2l_3^2\cos(2\phi_{23})\over2} \cos(\phi_{12}+\phi_{13})
.\ee
Therefore, \bea B_{123}^{\lle} &=& {- l_2^4l_3^4
\cos(2\phi_{23})\cos(\phi_{12}+\phi_{13}) \over 8}
 \int {d\chi\over \chi^6} W^2(\chi,\chi_s)
\int {d\chi'\over\chi'^6} W(\chi',\chi) W(\chi',\chi_s)   \vs
&&\times\left[ P_\Phi(l_2/\chi;\chi) P_\Phi(l_3/\chi';\chi')
+ P_\Phi(l_3/\chi;\chi) P_\Phi(l_2/\chi';\chi')
\right] + (\vl_1\leftrightarrow\vl_2) +
(\vl_1\leftrightarrow\vl_3).\eql{llbi}\eea

\subsection{Born Approximation}

The distortion tensor in \ec{psi} evaluates the potential everywhere along the
{\it unperturbed} path of the light. To go beyond the Born approximation, we
need to evaluate the potential at $\vx=\vx_0+\delta\vx$ where
\be
\delta x_a(\vt,\chi) =-\int d\chi' W(\chi',\chi) {\chi\over\chi'}
\Phi_{,a}(\vx_0;\chi')
.\ee
This leads to a new contribution to the distortion tensor, which in Fourier
space, reads
\be
\tilde\psi^{\born}_{ab}(\vl,\chi_s) = -\int{d\chi\over\chi^2} W(\chi,\chi_s)
\int {d\chi'\over \chi'^2} W(\chi',\chi)
\int {d^2l'\over (2\pi)^2}
l'_al'_bl'_c(\vl-\vl')_c \tilde\phi(\vl';\chi)\tilde\phi(\vl-\vl';\chi')
.\ee
This extra term in the distortion tensor contributes to the estimator for the
convergence
\be
{\hat{\tilde\kappa}}^{\born} ={-T_i(\vl)\over l^2}
\int{d\chi\over\chi^2} W(\chi,\chi_s)
\int {d\chi'\over \chi'^2} W(\chi',\chi)
\int {d^2l'\over (2\pi)^2}
F_i(\vl',\vl-\vl') \tilde\phi(\vl';\chi)\tilde\phi(\vl-\vl';\chi')
.\ee
Here
\be
F_i(\vl_1,\vl_2)\equiv 2\vl_1\cdot\vl_2 T_i(\vl_1)
.\ee
This expression is identical in form to that from lens-lens coupling, with the
substitution $E_i\rightarrow -F_i$. We can therefore copy the result from
\ec{billint} to get
\bea
B_{123}^{\born} &=& {-l_2^2l_3^2 T_i(\vl_1)\over 4l_1^2}
 \int {d\chi\over \chi^6} W^2(\chi,\chi_s)
\int {d\chi'\over\chi'^6} W(\chi',\chi) W(\chi',\chi_s)   \vs
&&\times\left[ P_\Phi(l_2/\chi;\chi) P_\Phi(l_3/\chi';\chi') F_i(\vl_2,\vl_3)
+ P_\Phi(l_3/\chi;\chi) P_\Phi(l_2/\chi';\chi')F_i(\vl_3,\vl_2)
\right] + (\vl_1\leftrightarrow\vl_2) +
(\vl_1\leftrightarrow\vl_3).\eea
But, $T_i(\vl_1)F_i(\vl_2,\vl_3)=l_1^2l_2^3l_3\cos(2\phi_{12})\cos(\phi_{23})$,
so
\bea
B_{123}^{\born} &=& {-l_2^3l_3^3 \cos(\phi_{23})\over 8}
 \int {d\chi\over \chi^6} W^2(\chi,\chi_s)
\int {d\chi'\over\chi'^6} W(\chi',\chi) W(\chi',\chi_s)   \vs
&\times&\left[ l_2^2 P_\Phi(l_2/\chi;\chi) P_\Phi(l_3/\chi';\chi')
\cos(2\phi_{12})
+ l_3^2 P_\Phi(l_3/\chi;\chi) P_\Phi(l_2/\chi';\chi')\cos(2\phi_{13})
\right] + (\vl_1\leftrightarrow\vl_2) +
(\vl_1\leftrightarrow\vl_3).\eea

\subsection{Summary}

Here we collect the results from the previous three subsection. The
reduced shear correction can be expressed in terms of the 2-point
function $P_2$:
 \bea B_{123}^{\rs} &=& {l_2^4l_3^4
P_2(l_2)P_2(l_3)\over 16}\left[ \cos(2\phi_{13}) +
\cos(2\phi_{12})\right]\vs &+& {l_1^4l_3^4 P_2(l_1)P_2(l_3)\over
16}\left[ \cos(2\phi_{23}) + \cos(2\phi_{12})\right]\vs &+&
{l_1^4l_2^4 P_2(l_2)P_2(l_1)\over 16}\left[ \cos(2\phi_{13}) +
\cos(2\phi_{23})\right]
 \eea

The other two corrections are best expressed in terms of \be
I(l_1,l_2) \equiv
 l_1^3l_2^3 \int {d\chi\over \chi^6} W^2(\chi,\chi_s)
\int {d\chi'\over \chi'^6} W(\chi',\chi) W(\chi',\chi_s)
 P_\Phi(l_1/\chi;\chi) P_\Phi(l_2/\chi';\chi') .\ee
Then,
the lens-lens term is
\bea B_{123}^{\lle} &=& {-l_2l_3
\cos(2\phi_{23})\cos(\phi_{12}+\phi_{13}) \over 8} \left[
I(l_2,l_3) + I(l_3,l_2)\right] \vs &-& {l_1l_3
\cos(2\phi_{13})\cos(\phi_{21}+\phi_{23}) \over 8} \left[
I(l_1,l_3) + I(l_3,l_1)\right] \vs &-& {l_2l_1
\cos(2\phi_{21})\cos(\phi_{32}+\phi_{31}) \over 8} \left[
I(l_2,l_1) + I(l_1,l_2)\right]. \eql{sumlens}\eea And the Born
term is
\begin{widetext}
\bea B_{123}^{\born} &=& {-\cos(\phi_{23})\over 8}
   \left[ l_2^2 I(l_2,l_3)\cos(2\phi_{12})
+ l_3^2 I(l_3,l_2) \cos(2\phi_{13}) \right] \vs &-&
{\cos(\phi_{13})\over 8}
   \left[ l_1^2 I(l_1,l_3)\cos(2\phi_{12})
+ l_3^2 I(l_3,l_1) \cos(2\phi_{23}) \right] \vs &-& {
\cos(\phi_{21})\over 8}
   \left[ l_2^2 I(l_2,l_1)\cos(2\phi_{32})
+ l_1^2 I(l_1,l_2) \cos(2\phi_{13}) \right] .\eql{sumborn}\eea
\end{widetext}

One word of caution: all of the above assume that our estimator for
$\kappa$ is as given in \ec{kest}. One could also imagine defining
the bispectrum as the three-point function of one-half of the trace
of the distortion tensor. These two expressions agree in the zeroth
order case, but they disagree when these higher order corrections
are included, because the distortion tensor is no longer the second
derivative of a potential $\psi$. Were we to be interested in the
latter definition, then the $\cos(2\phi)$ terms inside the square
brackets in \ec{sumborn} would be replaced by $1$. For the lens-lens
term, the product of cosines on the first line of \ec{sumlens} would
be replaced by $\cos^2\phi_{23}$; on the second line by
$\cos^2\phi_{13}$; and on the third by $\cos^2\phi_{12}$.
Practically, we think that the estimator of \ec{kest} is more
relevant, since it is the different components of ellipticity that
are measured, not the distortion tensor.

\section{How Important are the Corrections?}

There are a number of ways of assessing the importance of the
corrections considered here. First, we compute the skewness as a
function of smoothing angle. The smoothness at any given angle is
an integral over all configurations of the bispectrum with a
particular weighting scheme. Thus it reduces all elements to a
single number. Second, we can compute the equilateral
configuration as a function of multipole moment $l$. Comparing the
signal with the anticipated noise allows us to see whether this
one configuration, in which all $l$'s are equal, is sensitive to
the corrections. Finally, we compute the anticipated bias on a
cosmological parameter from future measurements of the bispectrum
if these corrections are neglected. If this bias is very small,
smaller than the anticipated statistical error, then there is no
need to worry about the corrections.

\subsection{Skewness}

The convergence skewness is defined as
\begin{equation}
S_3\equiv\langle \bar{\kappa}^3\rangle/\langle {\bar \kappa}^2\rangle^2
\end{equation}
where $\bar{\kappa}$ is the convergence smoothed over certain window
function. In the weakly nonlinear regime where second order
perturbation theory applies, Bernardeau et al. \citep{Bernardeau97} showed that the
skewness does not depend on the density fluctuation amplitude
$\sigma_8$ but is very sensitive to the mean matter density  $\Omega_m$.
This behavior holds even in the
highly nonlinear regime \cite{Hui99}. So $S_3$ is
particularly useful to break the degeneracy of $\Omega_m$ and
$\sigma_8$ in the lensing power spectrum. Detections of
skewness have been  reported by several groups
\citep{Bernardeau02,Pen:2003vw}; future surveys such as
Canada-France-Hawaii Telescope Legacy Survey \cite{Zhang03} could
determine $\Omega_m$ to $10\%$ in this fashion.

Corrections to the lensing bispectrum affect the prediction of
skewness and thus bias
the constraints of cosmological parameters. We
quantify  corrections of reduced shear, lens-lens coupling and
deviation from Born approximation to
$\langle \bar{\kappa}^3 \rangle$. $\langle \bar{\kappa}^3 \rangle$ is related to
 the lensing bispectrum by
\begin{equation}
\langle \bar{\kappa}^3 \rangle=\int B({\bf l}_1,{\bf l}_2,{\bf l}_3)W(l_1)W(l_2)
W(l_3)
\frac{d^2l_1}{(2\pi)^2}\frac{d^2l_2}{(2\pi)^2}\ .
\end{equation}
Here, $W(l)$ is the Fourier transform of the window function
$W(\theta)$. We study two window functions, the compensated Gaussian
and the aperture: \bea W_{\rm CG}(\theta)&=&(1-\theta^2/2\theta_f^2)
  \exp(-\theta^2/2\theta^2_f)
  \vs
  W_{\rm aperture}(\theta) &=&
  (1-\theta^2/\theta_f^2)(1/3-\theta^2/\theta_f^2)\Theta(\theta_f-\theta)
  \eea
  where $\theta_f$ is the characteristic
  scale in both cases. For both, $\langle \bar{\kappa}^3 \rangle$ is then a
  function of $\theta_f$. The corrections from the three effects
  considered in \S{III}
  and the total are shown in Fig.~\rf{skewness1}; on interesting scales,
  corrections are smaller than
  about $2\%$. Since
  $S_3\propto \Omega_m^{-\alpha}$ where $\alpha\sim 0.8$
  \cite{Bernardeau97,Zhang03}, these corrections could bias the
  determination of $\Omega_m$ by less than about $2\%$. So they  can
  be safely neglected in the near future.

\Sfig{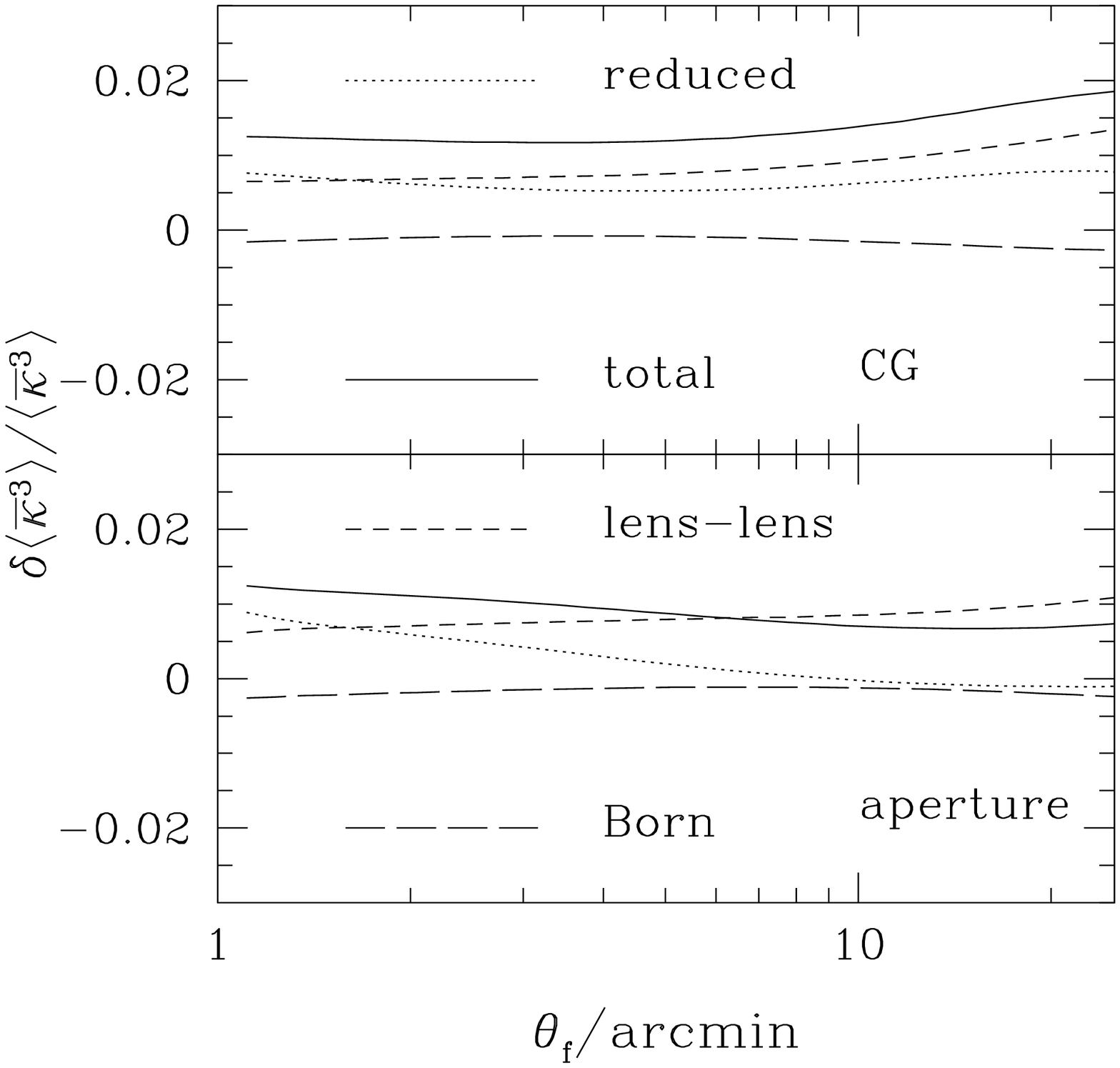}{Corrections to lensing skewness. The window functions we
  adopt are compensated Gaussian (top panel) and aperture function (bottom panel).
  Individual and combined corrections are less than
  about $2\%$ on all scales of interest.}

\subsection{Equilateral Configuration}

One configuration which is often used as a standard is the
equilateral configuration, wherein $l_1=l_2=l_3\equiv l$. The
bispectrum can then be plotted as a function of $l$. Let's consider
the corrections to the equilateral bispectrum.

 For the reduced shear
correction, all the cosines in \ec{rsbi} are $-0.5$, so adding up
all the permutations leads to \be B_{lll}^{\rs} =  {-3 l^8
P_2^2(l)\over 16} .\ee For the lens-lens correction, in the first
line of \ec{sumlens} $\phi_{12}+\phi_{13}=2\pi$;
$\cos(2\phi_{23})=-1/2$; and all three permutations contribute
equally, so \be
B_{lll}^{\lle} = {3l^2 \over 8} I(l,l)
.\ee
In the equilateral case of \ec{sumborn}, all the cosines are equal to $-1/2$, so
\be
B^{\born}_{lll}=-B^{\lle}_{lll}/2.\ee
The resulting corrections are shown in Fig.~\rf{blll}

\Sfig{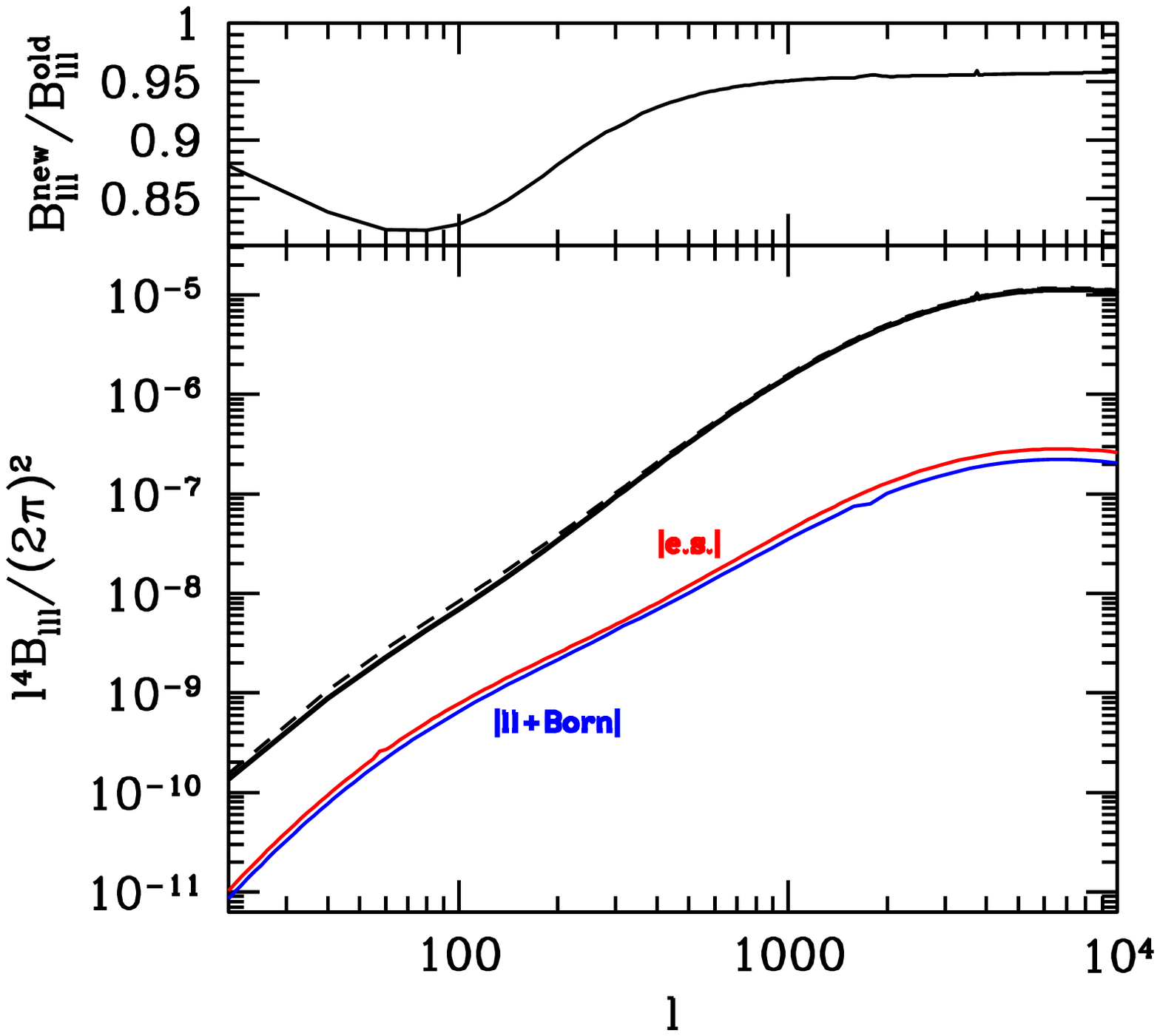}{The bispectrum of the convergence estimator for the
equilateral triangle configuration: $l_1=l_2=l_3\equiv l$. Thin
dashed line represents old calculation, which neglected second order
terms; thick line includes all relevant terms. Red curve is term
induced by second order relation between ellipticity and shear:
$\epsilon\sim 2\gamma\kappa$. Blue curve is the sum of the
contributions of lens-lens coupling and the Born correction. Note
that all second order corrections are negative. Top panel shows
fractional difference.}

The new terms are only about ten percent of the
first order term usually considered in this equilateral configuration.
Nonetheless, they may still be important
for precision cosmology where we sum over many different configurations.

\subsection{Cosmological Parameter Bias}

There is a simple formula relating the error in a cosmological
parameter to a mis-estimate in the theoretical prediction. \be
\Delta p = F^{-1} \sum_{l_1,l_2,l_3} w(l_1,l_2,l_3) {\partial
B_{l_1l_2l_3}\over \partial p } \Delta B_{l_1,l_2,l_3}
 .\eql{bias}\ee
 Here
the bias in the parameter is $\Delta p$; $F$ is the Fisher matrix
(here just one number since we treat the simple case of only one
parameter); $w$ is the weight, or the inverse variance, from the
experiment of interest, and $\Delta B$ is the mis-estimate in the
bispectrum, here taken to be the full set of corrections computed
above. The Fisher matrix too depends on the survey. It is \be F=
\sum_{l_1,l_2,l_3} w(l_1,l_2,l_3) \left({\partial
B_{l_1l_2l_3}\over\partial p}\right)^2 .\ee

The weights for a particular configuration depend on the survey in
question. We are interested in the question of whether these
corrections can ever be important, so we take the minimum possible
errors: cosmic variance due to simple Gaussian fluctuations.
Following Takada and Jain~\cite{Takada:2003ef}, the weights are
\be w(l_1,l_2,l_3)^{-1} = \Delta_{123} C_{l_1} C_{l_2}
C_{l_3}  \ee where the $C_l$'s are $l^4P_2(l)/4$, and
$\Delta=1$ when all $l$'s are different, $\Delta=2$ when two $l$'s
are equal, and $6$ when all $l$'s are equal. Under this weighting,
the sum extends over $l_1\le l_2\le l_3$.

A simple application of this formula is to consider the parameter to be
the amplitude $A$ of the bispectrum assuming the shape is known. Then, \ec{bias}
reduces to
\be
\left({\Delta A\over A}\right)_{\rm bias} = {\sum_{l_1,l_2,l_3} w(l_1,l_2,l_3)
B_{l_1l_2l_3} \Delta B_{l_1,l_2,l_3} \over
\sum_{l_1,l_2,l_3} w(l_1,l_2,l_3)
B_{l_1l_2l_3}^2 }
.\ee
This is to be compared with the fractional statistical error,
\be
 \left({\Delta A\over A}\right)_{\rm statistical} = \left[
\sum_{l_1,l_2,l_3} w(l_1,l_2,l_3) B_{l_1l_2l_3}^2 \right]^{-1/2}
.\ee

Before evaluating these sums, we can estimate them. The bias is of
order $\Delta B/B$ which, when summed over many modes, led to
percent level changes in skewness. Although the weighting scheme is
different here, we might still expect $\Delta A/A\sim 1\%$. The
statistical error is of order $\delta B_{\rm cos.\ var.}/B$ for each
configuration where both the cosmic variance error and the
bispectrum are plotted in \rf{zero}. The ratio is seen to be about
$10^3$. This is reduced by the square root of all configurations of
order $[l_{\rm max}^3]^{1/2}\sim 10^6$. Thus we expect the
statistical error to be of order $10^{-3}$. Evaluating, we find
\bea
\left({\Delta A\over A}\right)_{\rm bias} &=& 1.08\times 10^{-2} \vs
 \left({\Delta A\over A}\right)_{\rm statistical} &=& 1.22\times 10^{-3}
 \eea
for an all-sky survey, in agreement with our estimates. The
statistical error scales as $f_{\rm sky}^{-1/2}$, so we expect it to
be smaller than the bias for surveys that cover areas larger than
$f_{\rm sky}=0.01$, or $400$ square degrees.

So at least in principle, these corrections will eventually have to
be included. Ignoring them would induce a bias to the cosmological
parameters up to ten times larger than the anticipated statistical
error. Presently, of course, we are nowhere near these limits, so we
can safely neglect the corrections considered here when analyzing
lensing surveys.

\section{Conclusions}

We have computed corrections to the bispectrum due to: reduced
shear, lens-lens coupling, and the Born correction. These
corrections are smaller than the canonical term; this stems from the
fact that the spatial bispectrum is larger than the square of the
power spectrum. While the corrections are small and can be neglected
in present surveys, when areas as large as $400$ square degrees come
online, cosmological parameters extracted from the bispectrum will
be mis-estimated unless the corrections are included.

 This work is supported by the DOE and by NASA grant NAG5-10842.

\bibliography{v3}
\end{document}